\documentclass[aps,pra,twocolumn,amsmath,amssymb,nofootinbib,showpacs,superscriptaddress]{revtex4-1}
\usepackage[english]{babel}
\usepackage{latexsym}
\usepackage{graphics}
\usepackage{graphicx}
\usepackage{epsfig}
\usepackage{color}
\usepackage{bm}
\usepackage{amsmath}
\usepackage{amssymb}
\usepackage{amsthm}
\usepackage{dcolumn}
\usepackage{bm}
\usepackage{float}
\usepackage{hyperref}
\usepackage{color}
\usepackage{epstopdf}
\usepackage{braket}
\usepackage{cleveref}
\usepackage[svgnames]{xcolor}
\hypersetup{hidelinks,colorlinks=true,allcolors=DarkBlue}

\theoremstyle{remark}

\begin{document}

\preprint{APS/123-QED}

\title{Density-wave-type supersolid of two-dimensional tilted dipolar bosons}

\author{A.N. Aleksandrova}
\affiliation{Moscow Pedagogical State University, Moscow 119991, Russia}

\author{I.L. Kurbakov}
\affiliation{Institute of Spectroscopy, Russian Academy of Sciences, Troitsk, Moscow 142190, Russia}

\author{A.K. Fedorov}
\email{akf@rqc.ru}
\affiliation{Russian Quantum Center, Skolkovo, Moscow 121205, Russia}
\affiliation{National University of Science and Technology MISIS, Moscow 119049, Russia}

\author{Yu.E. Lozovik}
\email{lozovik@isan.troitsk.ru}
\affiliation{Institute of Spectroscopy, Russian Academy of Sciences, Troitsk, Moscow 142190, Russia}
\affiliation{Russian Quantum Center, Skolkovo, Moscow 121205, Russia}
\affiliation{National Research University Higher School of Economics, Moscow 109028, Russia}

\date{\today}
\begin{abstract}
We predict a stable density-waves-type supersolid phase of a dilute gas of tilted dipolar bosons in a two-dimensional (2D) geometry.
This many-body phase is manifested by the formation of the stripe pattern and elasticity coexisting together with the Bose-Einstein condensation and superfluidity at zero temperature.
With the increasing the tilting angle the type of the gas--supersolid transition changes from the first order to the second one despite the 2D character of the system,
whereas the anisotropy and many-body stabilizing interactions play crucial role.
Our approach is based on the numerical analysis of the phase diagram using the simulated annealing method for a free-energy functional. 
The predicted supersolid effect can be realized in a variety of experimental setups ranging from excitons in heterostructures to cold atoms and polar molecules in optical potentials.
\end{abstract}

\maketitle

\section{Introduction}

Remarkable progress on creating ultracold clouds of diatomic polar molecules~\cite{Ye2009,Dulieu2009,Ye2017}, 
degenerate gases of large-spin atoms~\cite{Pfau2005,Lev2011,Lev2012,Ferlaino2012}, 
and long-lived excitons in solid-state systems~\cite{Timofeev2006,Snoke2011,Butov2012,Rapaport2013,Geim2013,Novoselov2014,Alloing2014} 
makes it realistic to observe a large variety of interesting phenomena in dipolar systems 
and confirm seminal theoretical predictions~\cite{Lozovik1975,Lozovik1975,Lozovik1978,Lozovik1989,You1998,Shlyapnikov2000} (for a review, see Refs.~\cite{Baranov2008,Pfau2009,Baranov2012,Lozovik2019,Ferlaino20223}). 
Among non-conventional many-body phases of ultracold matter, a supersolid state attracts a special attention~\cite{Balibar2010,Svistunov2011,Prokofev2012,Balibar2012,Yukalov2020,Recati2023}. 
In such an unusual state, the condensate wavefunction has a lattice structure on top of a uniform background~\cite{Gross1957,Andreev1971,Nepomnyashchii1971,Rica1994,Shlyapnikov2015}. 
In addition to ultracold dipolar gases, supersolidity takes place in a range of systems, such as two-component systems~\cite{Saito2009}, Bose-Fermi mixtures~\cite{Buchler2003}, and condensates in optical lattices~\cite{Yi2007,Danshita2009}.  

There are several mechanisms for the realization of supersolidity in ultracold quantum dipolar gases.
Dilute weakly-interacting dipolar gases of bosons in two dimensions (2D) may demonstrate the roton-maxon structure of the spectrum by fine-tuning the short-range part of the interaction potential~\cite{Shlyapnikov2003}.
It is then possible to achieve vanishing the roton gap, so-called the roton instability regime, where the system is unstable with respect to periodic modulations of the order parameter~\cite{Rica1994}. 
However, instead of forming a supersolid state when approaching the roton instability, the condensate depletion diverges~\cite{Fischer2006,Cooper2007,Shlyapnikov2013}.
One of the possible way to avoid this divergence has been suggested in the case of tilted dipoles with the anisotropy of the excitation spectrum~\cite{Fedorov2014}.
Alternatively, the supersolid state of dipolar bosons can be rallied in the dense (strongly-correlated) regime~\cite{Pupillo2010,Pohl2010,Lozovik2011,Lozovik2011,Boninsegni2011}, in which the corresponding crystal has one particle per lattice site~\cite{PupilloZoller2007,Lozovik2007}. In this regime the supersolid state is possible in the presence of thermodynamically nonequilibrium defects in crystals only~\cite{Ceperley2004}. 
The presence of this phenomenon has been confirmed numerically~\cite{Troyer2006,Svistunov2006,Troyer2007,Lozovik2010,Burovski2005} and in experiments~\cite{Rittner2006,Sasaki2006}.

\begin{figure}
\center{\includegraphics[width=0.8\linewidth]{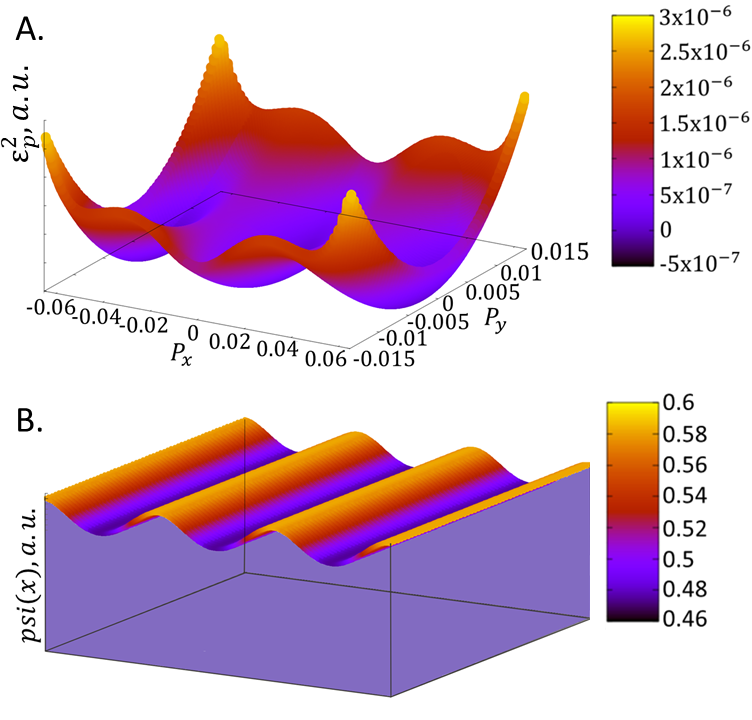}}
	\vskip-3mm
\caption{Supersolid state of a two-dimensional dilute gas of dipolar bosons.
In (a) the 2D anisotropic excitation spectrum of the system when the roton gap is vanishing, which leads to the instability of the homogeneous phase, is shown.
In (b) the 2D order parameter in the real space with the signatures of the density-wave-type supersolid is illustrated (the dimensionless parameter is $\alpha_3=0.402$ and is the tilting angle is $\theta=46^{\circ}$; for details, see Sec.~\ref{sec:Supersolid}).}
\label{fig:face}
\end{figure}

Nevertheless, one of the most intriguing questions relates to the possibility to obtain supersolidity in the dilute regime, 
where the mean-field description of the dilute trapped dipolar degenerate gases may be still valid~\cite{Fischer2006,Wilson2008,WilsonTicknor2012}.
The crucial requirement for obtaining supersolidity in the dilute regime is stabilizing the system~\cite{Petrov2015}, which can be achieved, for example, by adding a three-body repulsion~\cite{Shlyapnikov2015}. 
Without taking into account the many-body stabilization, the supersolid state can be considered as transient~\cite{Mishra2016}.
Recent advances in the study of supersolids of quantum gases are related to probing the roton excitation spectrum~\cite{Ferlaino2018,Ferlaino2019,Ferlaino2020,Ferlaino2021}, 
forming quantum stripes/droplets maintaining phase coherence~\cite{Pfau2017,Ferlaino20192,Modugno2019,Pfau2019} and eventually experimental observing supersolidity~\cite{Ferlaino2021}.
A complimentary mechanism for forming supersolidity in a dipolar quantum gas is related to the aforementioned anisotropy with respect to the rotational symmetry for a system of tilted dipoles,
which leads to the convergence of the condensate depletion up to the threshold of the roton instability~\cite{Fedorov2014}.
The latter approach to forming supersolid states has a potential advantage in the form of an additional level of controllability, which is possible by manipulating the tilting angle of dipoles.
We note that 2D systems of tilted dipoles and the formation of stripe phases have been studied via numerics yielding varying conclusions~\cite{Bombin2017, Cinti2019}. 

In this work, we solve the Gross-Pitaevskii (GP) equation by means of a straightforward minimization for the GP ({\it i.e.}, free-energy) functional. 
We consider a 2D system of tilted dipoles in a finite-thickness layer~\cite{Fedorov2014} under a stabilization by many-body effects~\cite{Petrov2015}. 
We predict a stable density-wave-type supersolid state of a 2D dilute Bose-Einstein condensed (BEC) gas of tilted dipoles. 
We obtain the full phase diagram of the system with both first and second kinds of the transition. 
We demonstrate the coexistence of superfluidity with elasticity and the crystalline stripe pattern at zero temperature (see Fig.~\ref{fig:face} for the illustration), 
which indicate on the density-wave-type supersolid phase, possible due to the smallness of the condensate depletion~\cite{Fedorov2014}.

Our paper is organized as follows.
In Sec.~\ref{sec:basics}, we introduce the model of tilted dipolar bosons in 2D and basic quantities of interest.
The technique that we use is based on the numerical minimization of the energy functional using stimulated annealing. 
In Sec.~\ref{sec:Supersolid}, we present the results of the numerical investigation of the model and observe indications of the supersolid phase of the density-wave type.
We summarize our results in Sec.~\ref{sec:conclusion}.  

\section{Tilted dipolar bosons: General relations}\label{sec:basics}

In this work, we consider a 2D BEC gas of tilted dipoles (for the detailed description of the system, see Ref.~\cite{Fedorov2014}).
We consider the corresponding free-energy functional in the following form:
\begin{equation}\label{F[psi]}
\begin{split}
	&F[\psi({\bf r})]-\mu N= \\ 
	&\!\!\!\!{\int}\left[\psi^*({\bf r})\left[\frac{(-i\hbar\nabla{+}m{\bf v})^2}{2m}{-}\mu\right]\psi({\bf r})+e_0(|\psi({\bf r})|^2)\right]d{\bf r} \\
	&+\frac12\int U({\bf r-s})|\psi({\bf r})\psi({\bf s})|^2d{\bf r}d{\bf s},
\end{split}
\end{equation}
which accounts for the 2D system (${\bf r}=\{x,y\}$ and
${\bf p}=\{p_x,p_y\}$) at zero temperature, $T=0$, 
while maintaining a constant chemical potential $\mu$. 
Here, $\psi({\bf r})$ represents the order parameter, $N=\int|\psi({\bf r})|^2d{\bf r}$ is the particle number, $m$ is the particle mass, and ${\bf v}$ is the velocity of the non-dissipative current.
The equation of state for the homogeneous phase with density $n=N/S$ is as follows:
\begin{equation}\label{e0n}
e_0(n)=\frac{g_2}{2} n^2+\frac{g_3}{6} n^3+\ldots
\end{equation}
Therefore, it describes the energy of the ground state per unit area as a function of the density $n$. 
The coupling constants are defined as follows:
\begin{equation}
\begin{split}
	&g_2=g_s+\left(3 \cos ^2 \theta-1\right)g_d,\quad g_s=\frac{2 \sqrt{2 \pi} \hbar^2}{m z_0} a_s,\\ 
	&\qquad\qquad\qquad\qquad g_d=\frac{2 \sqrt{2 \pi} \hbar^2}{m z_0} a_d. \\ 
\end{split}
\end{equation}
Here $g_2$ represents the two-body interaction, $g_s$ corresponds to the $s$-wave scattering, $g_d$ denotes the dipole-dipole interaction, and $g_3$ characterizes the three-body interaction. 
Additionally, $\theta$ stands for the tilting angle of dipoles with respect to the 2D plane, $z_0=\sqrt{\hbar/m\omega_z}$ is the oscillator length along the 2D plane direction, where $\omega_z$ is the corresponding oscillator frequency. 
The parameters $a_s$ and $a_d=md^2/3\hbar^2$ represent the 3D $s$-wave and dipole-dipole scattering lengths, respectively, with $d$ being the dipole moment of the particles;
$S=\int d{\bf r}\to\infty$ denotes the area of the periodic quantization box. 
In the Fourier transform  of the interparticle interaction $U({\bf r})$,
\begin{equation}
 	U({\bf r})=\int U({\bf p})e^{i{\bf pr}/\hbar}\frac{d{\bf p}}{(2\pi\hbar)^2},
\end{equation}
the momentum-dependent component in the following form has been distinctly isolated:
\begin{equation}
	U(\mathbf{p}) \equiv {\mathcal U}(\mathbf{p})-{\cal U}(0)=\int {\cal U}(\mathbf{r})\left(e^{-i \mathbf{p r} / \hbar}-1\right) d \mathbf{r}.
\end{equation}
Here ${\cal U}({\bf p})\!=\!\int {\cal U}({\bf r})e^{-i{\bf pr}/\hbar}d{\bf r}$ is the effective 2D interaction potential for the thin-layer motion~\cite{Fedorov2014}, and normalization condition $ U(0)\!=\!0$ being imposed. 
This normalization arises from the fact that the momentum-independent contribution $U(0)\!\!=\!\!\int U({\bf r})d{\bf r}\!=\!0$ has already been accounted for within the quantity $d^2e_0(n)/dn^2$. 
For tilted dipoles we have (see Ref.~\cite{Fedorov2014}):
\begin{equation}\label{Up}
	U({\bf p})=
	U_h({\bf p})\sin^2\theta+ U_v({\bf p})\cos^2\theta
\end{equation}
\begin{equation}\label{Uhp}
	U_h({\bf p})=\frac{4d^2}{\hbar}\int_0^{\infty}
	\frac{p_x^2dp_z}{p_x^2+p_y^2+p_z^2}
	\exp\left(-\frac{p_z^2z_0^2}{2\hbar^2}\right),
\end{equation}
\begin{equation}\label{Uvp}
	U_v({\bf p})=-\frac{4d^2}{\hbar}\int_0^{\infty}
	\frac{(p_x^2+p_y^2)dp_z}{p_x^2+p_y^2+p_z^2}
	\exp\left(-\frac{p_z^2z_0^2}{2\hbar^2}\right),
\end{equation}
and, finally, we assume the two-dimensionality of the problem
($\hbar\omega_z\gg4\pi\hbar^2n/m$) and the weakly interacting regime ($g_s,g_d,g_3n\ll4\pi\hbar^2/m$).

\subsection{Trial wavefunction and details of the minimization}

The Bogoliubov excitation spectrum,
\begin{equation}\label{eBp}
\begin{split}
	&\varepsilon_{\bf p}=\sqrt{T_{\bf p}(T_{\bf p}+2U_{\bf p})}, \\ 
	&T_{\bf p}\equiv\frac{p^2}{2m},\; \\
	&U_{\bf p}\equiv\left(\frac{d^2e_0(n)}{dn^2}+ U({\bf p})\right)n,
\end{split}
\end{equation}
of systems with the free-energy functional of the form given by Eq.~(\ref{F[psi]}) exhibit a roton-maxon effect along the $y$ axis, as expressed by Eqs.~(\ref{Up})--(\ref{Uvp}), 
which is stronger than along the $x$ axis, i.e., $\varepsilon_{p,0}>\varepsilon_{0,p}$. 
As a result, the density wave (DW) in the emerging supersolid phase may be oriented along the $x$ axis, with the vector aligned along the $y$ axis. 
Consequently, as a trial function for the functional $F[\psi({\bf r})]-\mu N$ [see Eq.~(\ref{F[psi]})], 
we consider complex-valued functions that are periodic with a period $\lambda$, depending only on the variable $y$:
\begin{equation}\label{psir}
	\psi({\bf r})\equiv\psi(y)=\psi(y+\lambda).
\end{equation}
In the Fourier series expansion of functions $\psi(y)$, we consider a finite number of harmonics to achieve the desired accuracy. 
The function $e(|\psi(y)|^2)$ is integrated in the position representation. 
For the undeformed DW, we are looking for the global minimum of the functional with respect to both $\lambda$ and the amplitudes of the harmonics. 
In the case of the deformed DW, the minimization is performed only with respect to the amplitudes of the harmonics. 
In the absence of the velocity, the function $\psi(y)$ is an even real function.

The ground state of a supersolid, which is characterized by a stationary and undeformed DW, is to be determined by minimizing the functional $F-\mu N$ with a fixed chemical potential $\mu$ [see Eq.~(\ref{F[psi]})], 
according to principles of a first-order phase transition. As a result of this procedure, we obtain $\psi_0(y)$, with which we calculate the particle number as $N = \int|\psi_0(y)|^2dy$ for the same $\mu$. 
Subsequently, while keeping the particle number fixed at $N$~\cite{Fisher1973}, we minimize the functional $F$ with additional constraints imposed on the velocity $dv$ and deformation $d\beta$. 
From this minimization, we derive the tensors for helicity modulus and deformation.

Below we show the results of the numerical minimization of both functionals, $F$ and $F-\mu N$, with the use of the simulated annealing method. 
To compute the ground state of the supersolid, we start with a random configuration and set the initial ``temperature'' $\mathcal{T}$ in the Metropolis algorithm to be of the order of $Se_0(n)$. 
For the calculation of the deformation tensor and the superfluid component, we use the previously determined ground state profile as the initial configuration. 
The initial ``temperature'' is on the order of $Nm(d{\bf v})^2/2$ when calculating the helicity modulus tensor and $Se_0(n)(d{\bf a})^2/2$ when computing the deformation tensor. 
The final ``temperature'' is approximately 12--16 orders of magnitude lower than the initial ``temperature'', and the ``temperature'' reduction follows a monotonically decreasing exponential trend. 
The number of iterations is chosen to achieve complete annealing, and poorly annealed calculations in the immediate vicinity of phase transitions are disregarded. 
For a fine exploration of the first-order phase transition, we employ multiple annealings, initiating from various random configurations.

\subsection{Computed quantities}

For both phases, homogeneous gas and the supersolid DW, we numerically compute a number of quantities. 
\begin{itemize}
\item The compressibility
\begin{equation}\label{mes-chi}
	\frac{m^2}{\chi}=\frac1{\partial n/\partial\mu}.
\end{equation}

\item The pressure
\begin{equation}\label{mes-P}
	P=\mu n-\frac{F_0}S.
\end{equation}
Here, $F_0=F[\psi_0(y)]$ represents the value of the functional $F$, when $F-\mu N$ reaches its minimum, and $\psi_0(y)$ is the value of the order parameter at the minimum of $F-\mu N$ at a fixed $\mu$. 
In the gas phase $\psi_0(y)=\sqrt{n}$.

\item The square of the roton gap in the Bogoliubov spectrum (\ref{eBp}) in the homogeneous phase is given by
\begin{equation}\label{mes-Er2}
	E_{\rm r}^2=\min_{\bf p}\varepsilon_{\bf p}^2,
\vspace{-1mm}
\end{equation}
with the minimization performed separately for $p_x$ and $p_y$, starting from the maxon momentum (if it exists).

\item The magnitude of the diagonal long-range order (DLRO) for the DW is given by
\begin{equation}\label{mes-Gamma}
	\Gamma=\int_0^{\lambda_0}\frac{(|\psi_0(y)|^2-n)^2}{n^2}\frac{dy}{\lambda_0},
\end{equation}
where $\lambda_0$ is the value of the DW period $\lambda$ at the minimum of $F-\mu N$.

\item The diagonal elements $Y_x$ and $Y_y$ of the helicity modulus tensor for the superfluid component ~\cite{Lozovik2017} are given by
\begin{equation}\label{mes-YxYy}
\begin{split}
	&Y_x=\left.\frac1{m^2S}\frac{d^2F_0({\bf v})}{dv_x^2}\right|_{{\bf v}=0}, \\
	&Y_y=\left.\frac1{m^2S}\frac{d^2F_0({\bf v})}{dv_y^2}\right|_{{\bf v}=0}.
\end{split}
\end{equation}
Here $F_0({\bf v})$ represents the value of $F_0$ with fixed ${\bf v}$.

The temperature $T_c$ of the Berezinskii-Kosterlitz-Thouless transition~\cite{Kosterlitz1973} (crossover~\cite{Lozovik20072}) is determined by the total superfluid density $n_s(T)$ as $T_c=\pi\hbar^2n_s(T_c)/2m$.  
The temperature-dependent quantity $n_s\equiv n_s(T)$ can be calculated as the {\it geometric} mean~\cite{Minnhagen1991,You2012} of $x$ and $y$ components of the helicity modulus 
\begin{equation}\label{mes-ns}
\frac{n_s}{m} = \sqrt{Y_x Y_y}.
\end{equation}

\item The stretching-compression deformation coefficient $u_x$ and the shear deformation coefficient $u_y$ (which, for the DW, is equivalent to the rotation of the DW) are given by
\begin{equation}\label{mes-uxuy}
\begin{split}
&u_x=\left.\frac1S\frac{d^2F_0((1+\beta)\lambda_0)}
{d\beta^2}\right|_{\beta=0},\\
&u_y=\left.\frac1S\frac{d^2F[\psi({\rm R}^{\beta}{\bf r})]}
{d\beta^2}\right|_{\beta=0}.
\end{split}
\end{equation}
Here, $F_0(\lambda)$ represents the value of $F_0$ with fixed $\lambda$, and
\begin{equation}\label{R(beta)}
\mathrm{R}^\beta=\left(\begin{array}{cc}
\cos (\beta) & \sin (\beta) \\
-\sin (\beta) & \cos (\beta)
\end{array}\right)
\end{equation}
is the rotation matrix by an angle $\beta$. 
For the convenience, we rotate the Hamiltonian with respect to the DW, using the function $U({\rm R}^{-\beta}{\bf p})$ instead of $U({\bf p})$.
\end{itemize}

\section{Supersolid density wave}\label{sec:Supersolid}

The results presented below numerically indicate the existence of a stable supersolid DW following the scenario of roton-like attraction with stabilizing many-body repulsion~\cite{Shlyapnikov2015} and the decisive role of anisotropy~\cite{Fedorov2014}.
The calculations are performed for $m=164$ a.u. and $a_d=7$ nm, which correspond to dysprosium atoms~\cite{Lev2011}.
The problem is characterized by five dimensionless control parameters
\begin{equation}
	\nu=\frac{a_d}{z_0}, \; \eta=\frac{\mu m z_0^2}{\hbar^2}, \; \alpha=\frac{g_s}{g_d}, \; \alpha_3=\frac{m g_3}{2 \pi \hbar^2 z_0 a_d}
\end{equation}
and $\theta$. In all the data, we use the following set of parameters: 
$\nu=7/150$, $\eta=0.0042$, and $\alpha=-3/7$, corresponding to $\mu=0.76$ nK, $z_0=150$ nm, $a_s=-3$ nm, and work within the variables $\alpha_3$ and $\theta$. 

\begin{figure}[t]
\centering\includegraphics[width=8cm]{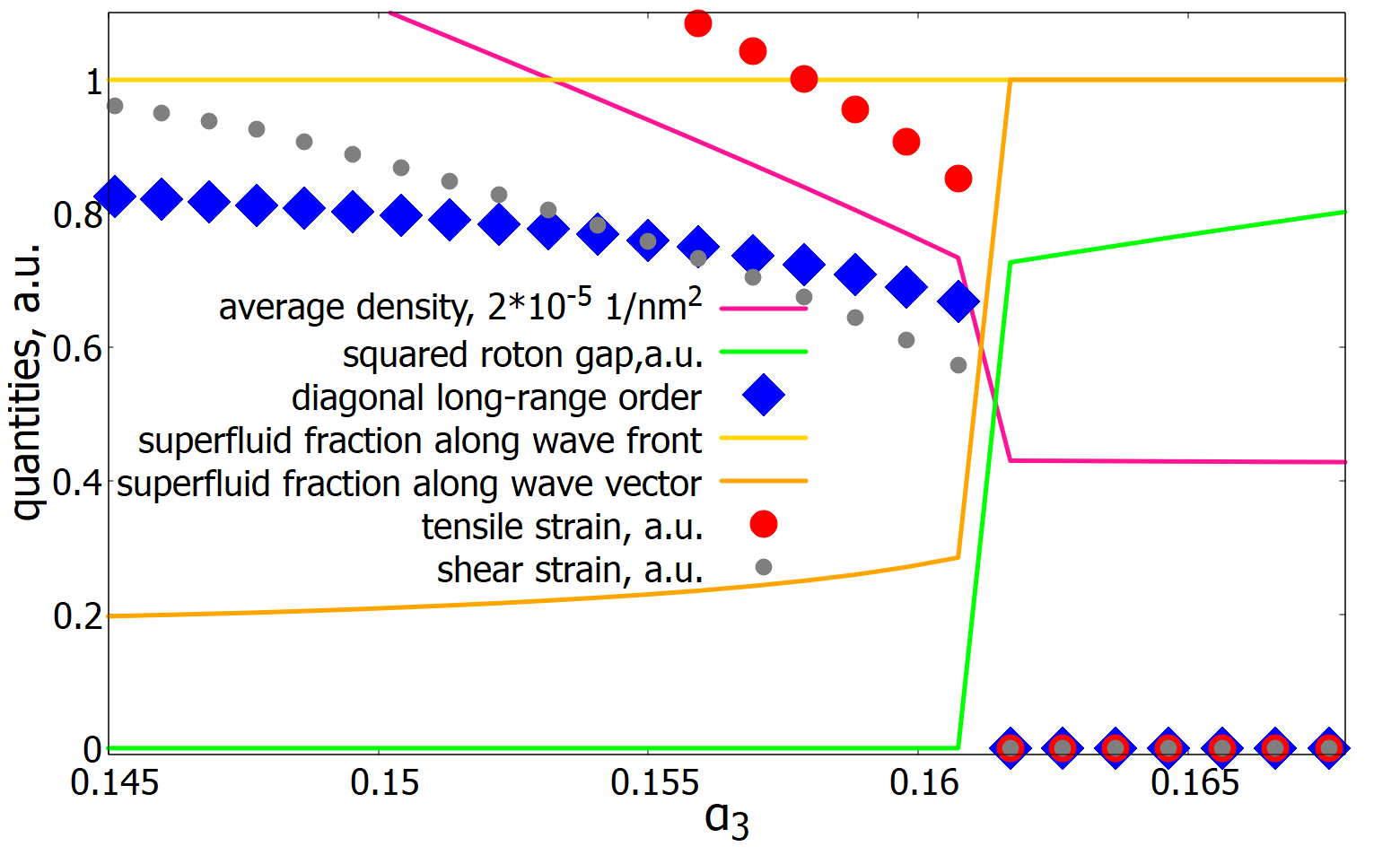}
\centering\includegraphics[width=8cm]{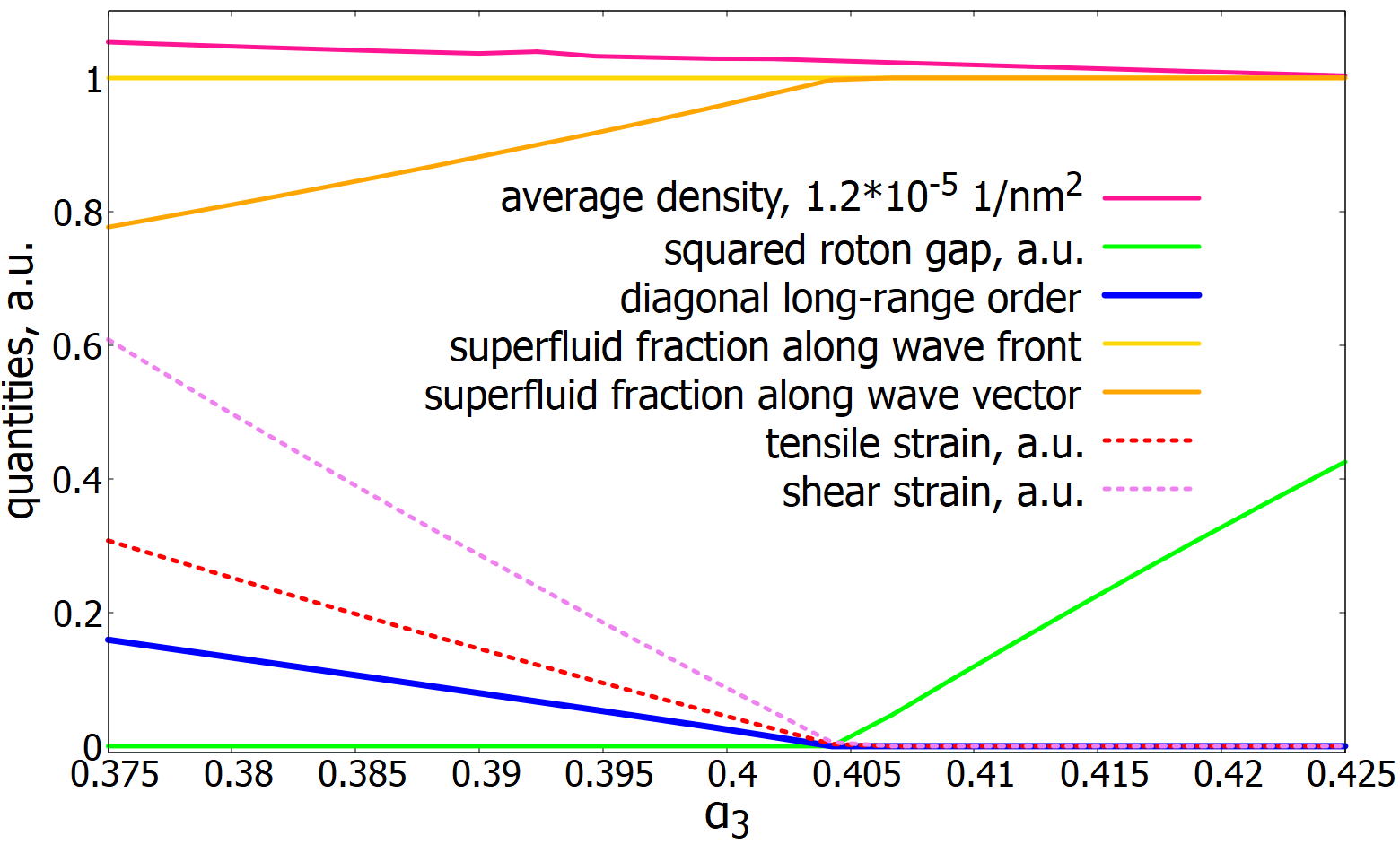}
\vskip-4mm
\caption{(a) First-order gas to supersolid density wave transition at $\theta=44.3^{\circ}$ and (b) second-order transition at $\theta=46^{\circ}$ are illustrated. 
Both components of the helicity modulus tensor, $Y_x$ and $Y_y$, and the deformation tensor, $u_x$ and $u_y$, are shown, 
along with the diagonal long-range order parameter $\Gamma$, the square of the roton gap $E_{\rm r}^2$, and the average density $n$. 
The first (second) order transition corresponds to the jump in (a) (kink in (b)).}
\label{firstsecond}
\vskip-4mm
\end{figure}

\subsection{Stable supersolid density wave}

According to the numerical calculation with $\alpha_3=0.161$ and $\theta=44.3^{\circ}$, 
we clearly observe the effect of a supersolid density wave in a 2D array of tilted dipoles. 
The squared profile of the order parameter $|\psi_0(y)|^2$ exhibits periodic oscillations with a relative amplitude of $(n_{\rm max}-n_{\rm min})/n=3.2$. 
The magnitude of DLRO [see Eq.~(\ref{mes-Gamma})] is $\Gamma=1.26$, which indicates on its non-zero value.
Superfluidity is also observed. Both diagonal elements of the helicity modulus tensor in its principal axes, $Y_x=n/m$ and $Y_y=0.28n/m$, are non-zero.
Elasticity is present since the stretching-compression deformation coefficient $u_x=1.4e_0(n)$m and the shear deformation coefficient $u_y=2.9e_0(n)$ are both non-zero.
Both stability-related quantities, the compressibility $m^2/\chi=0.0053(4\pi\hbar^2/m)$ and the pressure $P=10.2e_0(n)$, are positive.
Finally, in the 2D weakly correlated system at $T=0$, a Bose-Einstein condensate is indeed present, and its existence does not require further verification.
To conclude, we see all the features of the supersolid phase.

We summarize our results in phase diagrams, which are presented in Fig.~\ref{firstsecond}.
As is known, in the 2D isotropic case, the gas--supersolid transition is of the first order~\cite{Shlyapnikov2015}. 
Therefore, even with a sufficiently weak anisotropy, the nature of this transition should be of the first kind. 
This is precisely evident in Fig. \ref{firstsecond}(a): at the tilt angle $\theta=44.3^{\circ}$, 
the magnitude of the DLRO, both components of the strain and helicity modulus tensors, as well as the square of the roton gap and the average density, undergo abrupt changes at the transition point.

\begin{figure}[t]
\centering\includegraphics[width=8cm]{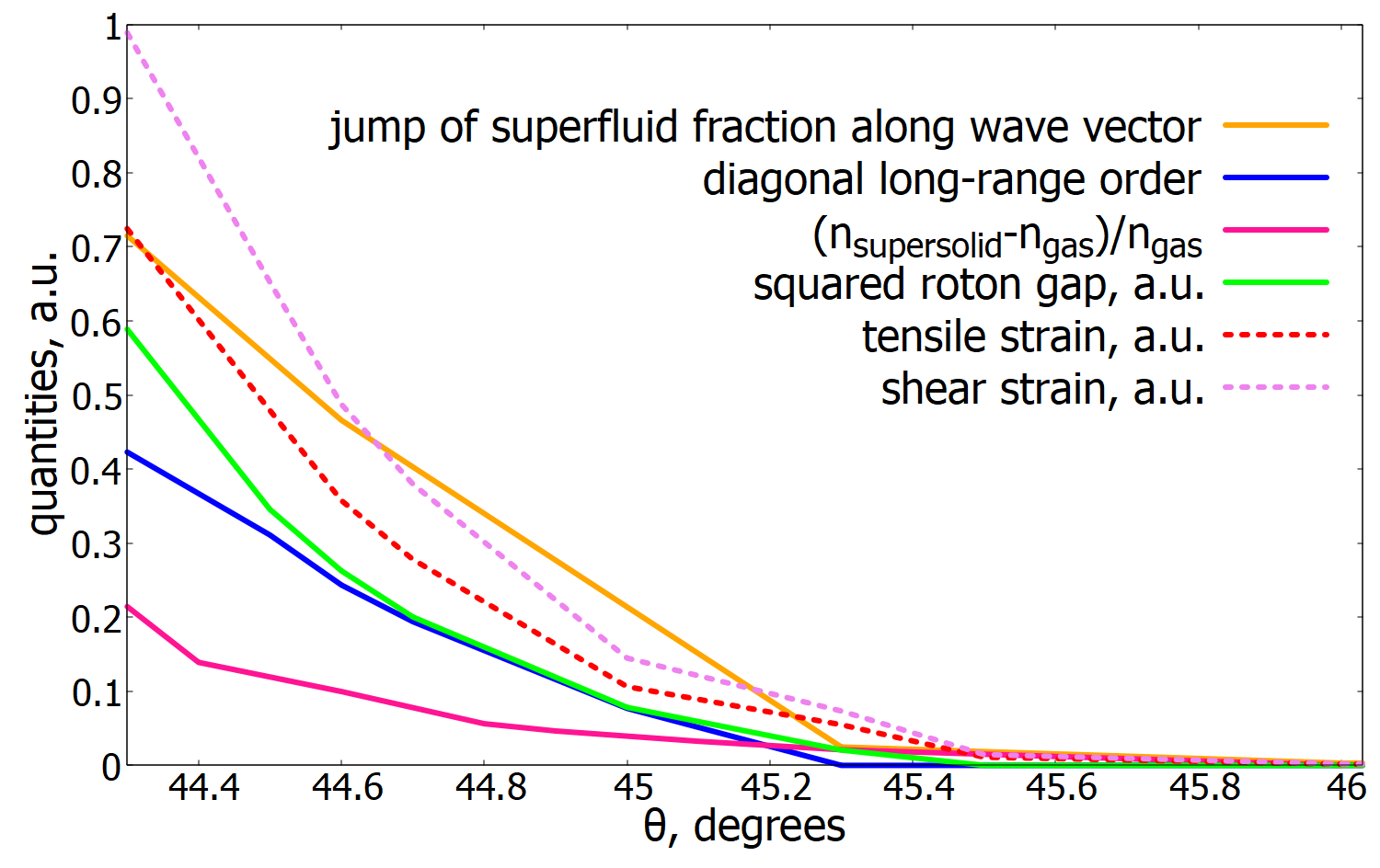}
\centering\includegraphics[width=8cm]{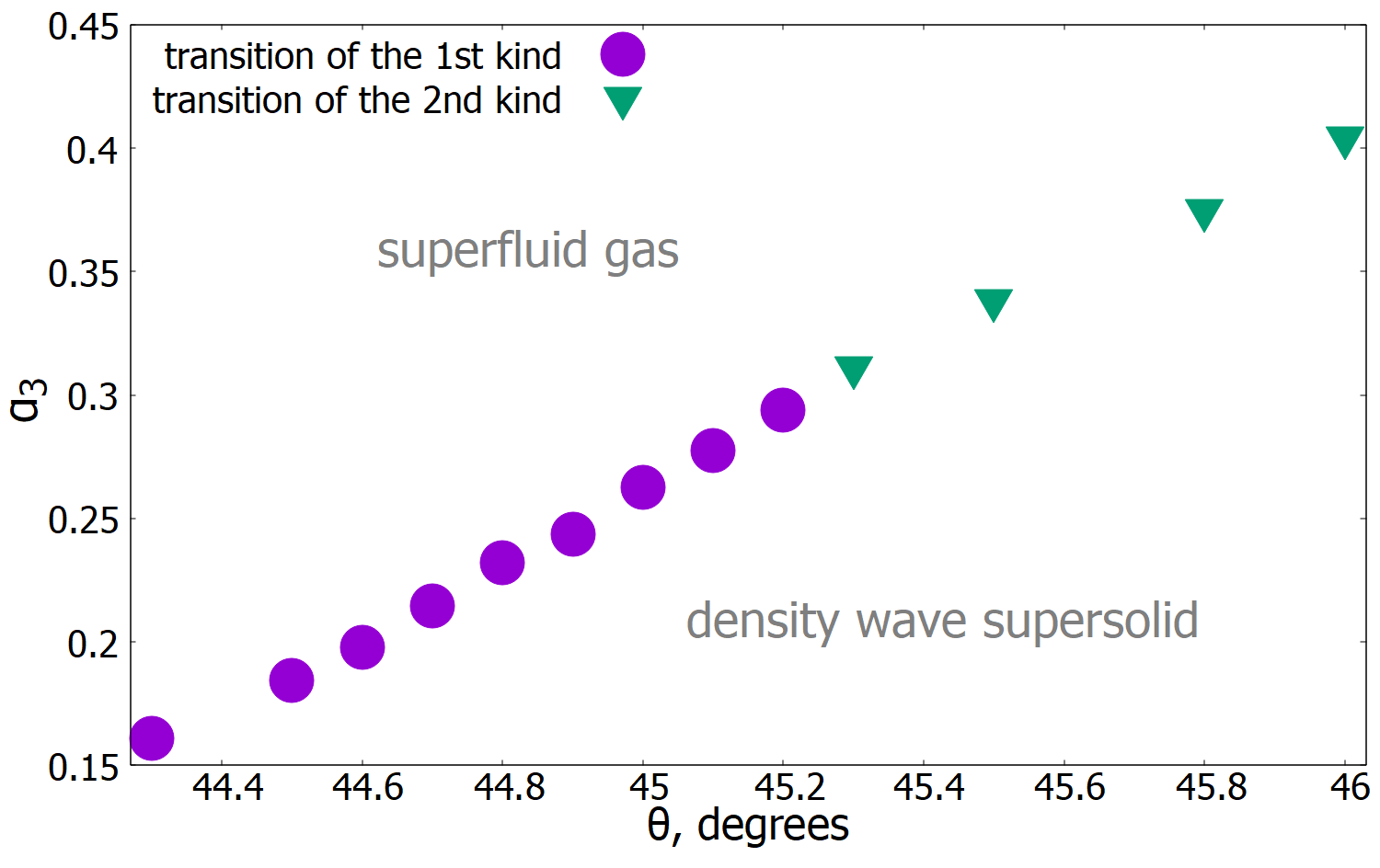}
\vskip-4mm
\caption{
In (a) the jump magnitudes at the transition are shown for the $Y_y$ component of the helicity modulus, both components $u_x$ and $u_y$ of the strain tensor, the DLRO magnitude $\Gamma$,
 the density jump between supersolid and gas, and the square of the roton gap $E_{\rm r}^2$, as functions of the tilt angle $\theta$. 
 (b) The phase diagram in the $\theta$--$\alpha_3$ variables is presented.}
\label{jumpspd}
\vskip-4mm
\end{figure}

However, with increasing anisotropy, i.e., with the growth of the tilt angle $\theta$, the magnitudes of these jumps {\it decrease}, as it is illustrated in Fig.~\ref{jumpspd}(a). 
At the certain critical value $\theta_c=45.3^{\circ}$, all jumps simultaneously {\it vanish}. 
With further increase in anisotropy at angles $\theta>\theta_c$, only {\it kinks} are present instead of jumps.
These kinks indicates on a second-order transition, as shown in Fig.~\ref{firstsecond}(b). 
The phase diagram of the gas--supersolid transition has a point of intersection between the first and second order transitions at $\theta=\theta_c$ 
[see Fig.~\ref{jumpspd}(b)]: the first order for weaker anisotropy ($\theta<\theta_c$) and the second order for stronger anisotropy ($\theta>\theta_c$).

\section{Discussion and Conclusion}\label{sec:conclusion}

In this work, we have presented the results of the investigation of the supersolidity effect in the system of tilted dipolar bosons with stabilizing many-body repulsion.
We have identified the second-order transition from gas to density wave supersolid and the point of intersection with the first-order transition line appear to be a general feature in 2D at $T=0$ and inherently arises from the system's anisotropy.
The observed pattern of first and second-order transitions in the gas-supersolid transition for anisotropic 2D systems aligns with the touching of the Bogoliubov spectrum zero-energy points at two roton momenta $\pm{\bf k}_{\rm roton}$~\cite{Fedorov2014}.
Indeed, the zero roton gap allows for the {\it macroscopic population transfer} from the homogeneous condensate to these two states due to the touching. 
Consequently, the superposition leads to the density wave of the condensate: $\psi({\bf r})=a_0+a_1\cos({\bf k_{\rm roton}r})+...$.
Furthermore, the absence of Fischer-like~\cite{Fischer2006} divergence beyond the condensate in the conditions of roton gap closing at two points~\cite{Fedorov2014} allows the system to remain within the regime of weak correlations.

As we expected, the proposed approach can be used in ongoing experiments on the study of supersolidity of dipolar bosons. 
In particular, in the experiment with untilted dipoles ($\theta=0$), but in a cigar-shaped trap~\cite{Biagioni2022}, a point of the intersection between first and second-order transitions in the gas--supersolid transition has also been observed 
(this is consistent with the theory provided in Ref.~\cite{Biagioni2022}). 
In this experiment, the anisotropy is imposed by the geometry: DWs are indeed formed along the cigar.
In our case, in the 2D system anisotropy can be imposed by the control of the electric field in the case of polar molecules~\cite{Ye2009,Dulieu2009,Ye2017}
or an external magnetic field for ultracold atoms with induced dipole moment, 
such as erbium~\cite{Ferlaino2012} or dysprosium~\cite{Lev2011,Lev2012}.

\section*{Acknowledgments}

The work of Yu.E.L. is supported by the Russian Science Foundation Grant No. 23-12-00115 (studying supersolid states). 
A.K.F. acknowledge the support by the Russian Science Foundation Grant No. 19-71-10092 (analysis of the phase diagram) and the Priority 2030 program at the National University of Science and Technology ``MISIS” under the project K1-2022-027.
The work was also supported by the Russian Roadmap on Quantum Computing (Contract No. 868-1.3-15/15-2021). 

\bibliography{bibliography-dipoles.bib, addition.bib}

\newpage

\end{document}